\documentclass[12pt]{article}

\usepackage{amssymb}
\usepackage{amsxtra}
\usepackage{amsmath}
\usepackage{amsfonts}
\usepackage{CJK}
\usepackage[titletoc]{appendix}
\usepackage{graphicx}
\usepackage{color}

\numberwithin{equation}{section}

\newtheorem{thm}{Theorem}[section]

\newtheorem{prop}[thm]{Proposition}

\newtheorem{ass}[thm]{Assumption}

\newtheorem{lem}[thm]{Lemma}

\newtheorem{rem}[thm]{Remark}

\newcommand{\eqa}{\begin{eqnarray}}
\newcommand{\eeqa}{\end{eqnarray}}
\newcommand{\beq}{\begin{equation}}
\newcommand{\eeq}{\end{equation}}
\newcommand{\nn}{\nonumber}

\allowdisplaybreaks

\begin{document}
\title{ {An initial-boundary value problem for the coupled focusing-defocusing complex short pulse equation with a $4\times4$ Lax pair}}

\author{{\footnotesize { Bei-bei Hu$^{1,2}$ , Tie-cheng Xia$^{1,}$\thanks{Corresponding author. E-mails: hubsquare@chzu.edu.cn(B.-b. Hu), xiatc@shu.edu.cn(T.-c. Xia), mawx@cas.usf.edu(W.-x. Ma)} , Wen-xiu Ma$^{3}$}}\\
{\footnotesize { {\it$^{1}$Department of Mathematics, Shanghai University, Shanghai 200444, China}}}\\
{\footnotesize { \it $^{2}$School of Mathematics and Finance, Chuzhou University, Anhui, 239000, China}}\\
{\footnotesize { \it $^{3}$Department of Mathematics and Statistics, University of South Florida, Tampa, FL, 33620-5700, USA
}}
}
\date{\small \today}\maketitle

\textbf{Abstract}: In this paper we investigate the coupled focusing-defocusing complex short pulse equation, which describe the propagation of ultra-short optical pulses in cubic nonlinear media. Through the unified transform method, the initial-boundary value problem for the coupled focusing-defocusing complex short pulse equation with $4\times 4$ Lax pair on the half-line are to be analyzed. Assuming that the solution $\{q_1(x,t),q_2(x,t)\}$ of the coupled focusing-defocusing complex short pulse equation exists, we show that $\{q_{1,x}(x,t),q_{2,x}(x,t)\}$ can be expressed in terms of the unique solution of a $4\times 4$ matrix Riemann-Hilbert problem formulated in the complex $\lambda$-plane. Thus, the solution $\{q_1(x,t),q_2(x,t)\}$ can be obtained by integration with respect to $x$. Moreover, we also get that some spectral functions are not independent and satisfy the so-called global relation.

\section{Introduction}

One of the most important integrable systems in mathematics and physics is the following short plus (SP) equation
\beq u_{xt}=u+\frac{1}{6}(u^3)_{xx}, \label{1.1}\eeq
where $u=u(x,t)$ is a real-valued function, representing the magnitude of the electric field, the subscripts $t$ and $x$ denote partial differentiation,
which can describe the propagation of ultra-short optical pulses in silica optical fibers. The SP equation was derived by Sch\"{a}fer and Wayne \cite{Sch2004}. Actually, the SP equation appeared first as one of Rabelo's equations which describe pseudospherical surfaces, possessing a zero-curvature
representation \cite{Rabelo1989}. The SP equation is completely integrable\cite{Sakovich2005}, and has been studied extensively on the bi-Hamiltonian structure\cite{Brunelli2006}, Cauchy problem \cite{Okamoto2017}, integrable self-adaptive moving mesh schemes \cite{Feng2014}, soliton solutions given by the Darboux transformation (DT) method\cite{Liu2017}. Besides, the long-time behavior for the solutions of decay initial value problem of SP equation has been analyzed by nonlinear steepest descent method/Deift-Zhou method\cite{XJ2016},
Moreover, Boutet and Monvel et al. have studied Riemann-Hilbert (RH) problem of the SP equation \cite{Monvel2017} by using the inverse scattering transform (IST) method.

However, the IST method can only be used to analyze pure initial value problems. In 1997, Fokas used IST thought to construct a new unified method, we call this method as Fokas method, this method can be used to study the initial-boundary value (IBV) problems for both linear and nonlinear integrable evolution PDEs with $2\times 2$ Lax pairs\cite{Fokas1997,Fokas2002,Fokas2005,Lenells2011,lenells2011,Zhang2017,Xia2017}.
Just like the IST on the line, the Fokas method yields an expression for the solution of an IBV problem in terms of the solution of a RH problem. In particular, an effective way analyzed the asymptotic behaviour of the solution is based on this RH problem and by employing the nonlinear version of the steepest descent method introduced by Deift and Zhou \cite{Deift1993}. In 2012, Lenells extended the Fokas method to study the IBV problems for integrable nonlinear evolution equations with $3\times 3$ Lax pairs \cite{Lenells2012}. After that, more and more researchers begin to pay attention to studying IBV problems for integrable evolution equations with higher-order Lax pairs, the IBV problem for the many integrable equations with $3\times 3$ or $4\times 4$ Lax pairs are studied, such as, the Degasperis-Procesi equation \cite{Lenells2013,Monvel2013}, the Ostrovsky-Vakhnenko equation \cite{Monvel2015}, the Sasa-Satsuma equation \cite{Xu2013}, the three wave equation \cite{Xu2014}, the spin-1 Gross-Pitaevskii equations \cite{Yan2017} and others \cite{Geng2015,Xu2016,Monvel2016,Liu2016,Tian2017}. These authors have also done some work about integrable equations with $2\times2$ or $3\times3$ Lax pairs \cite{Zhang2017,Hu1,Hu2,Hu3,Hu4}.

Similar to the nonlinear Schr\"{o}dinger(NLS) equation, it is known that the complex-valued function has advantages in describing optical waves which have both the amplitude and phase information, Following this spirit, Feng \cite{feng2015} proposed a complex short pulse (CSP) equation
\beq q_{xt}+q+\frac{1}{2}\epsilon(|q|^2q_x)_{x}=0, \label{1.2}\eeq
where, $\epsilon=\pm1$ represents focusing- and defocusing-type, $q(x,t)$ is the complex function. It can be derived from the Maxwell equation.
The CSP equation has been studied extensively on the integrability associated with explicit form of the Lax pair \cite{feng2015}, geometrically and algebraically \cite{Feng2017,Shen2016}, multiply smooth soliton, loop soliton, cuspon soliton, breather soliton and rogue wave solutions given by the DT method \cite{Feng2016,Zha2017}, periodic traveling wave solutions given by the $F$-expansion method \cite{Li2017},

To describe the propagation of optical pulses in birefringence fibers, Feng \cite{feng2015} also proposed a coupled complex short-pulse (CCSP) equation
\eqa \left\{ \begin{array}{l}
q_{1,xt}+q_1+\frac{1}{2}[(\epsilon_1|q_1|^2+\epsilon_2|q_2|^2)q_{1,x}]_{x}=0,\\
q_{2,xt}+q_2+\frac{1}{2}[(\epsilon_1|q_1|^2+\epsilon_2|q_2|^2)q_{2,x}]_{x}=0.
\end{array} \right. \label{1.3}\eeqa
Similarly, where $\epsilon_i=\pm1$ means focusing case and defocusing case, $\epsilon_1=-\epsilon_2=1$ means focusing-defocusing case, $q_1(x,t)$ and $q_2(x,t)$ are the complex functions of $x$ and $t$, which indicate the magnitudes of the electric fields. The CCSP equation has been studied extensively on the integrability associated with explicit form of the Lax pair, conservation laws \cite{feng2015}, bi-Hamiltonian structure, bilinearization \cite{Shen2016}, soliton solutions given by the Hirota's bilinear method \cite{feng2015,Guo2016,Yang2017}, rogue waves solutions given by the DT method \cite{Ling2016}.

Most recently, Yang and Zhu consider the following coupled focusing-defocusing complex short pulse equation \cite{Yang2017}
\eqa \left\{ \begin{array}{l}
q_{1,xt}+q_1+\frac{1}{2}[(|q_1|^2-|q_2|^2)q_{1,x}]_{x}=0,\\
q_{2,xt}+q_2+\frac{1}{2}[(|q_1|^2-|q_2|^2)q_{2,x}]_{x}=0.
\end{array} \right. \label{1.4}\eeqa
by Hirota's bilinear method, the bright-bright, bright-dark, dark-dark soliton solutions and rogue waves solutions to be constructed.
In this paper, we investigate the IBV problems for the following coupled focusing-defocusing complex short pulse equation on the half-line via a unified transform method. The IBV problems of system \eqref{1.4} on the interval are presented in anothers paper. For the focusing case ($\epsilon_1=\epsilon_2=1$) and defocusing case ($\epsilon_1=\epsilon_2=-1$) which can be study by the same ways.

Throughout this paper, we consider the half-line domain $\Omega$ and the IBV problems for the system \eqref{1.4} as follows
\eqa\begin{array}{l}
$Half-line domain (see Figure 1)$: \Omega=\{0<x<\infty,0<t<T\};\\
$Initial values$: u_0(x)=q_1(x,t=0),\; v_0(x)=q_2(x,t=0);\\
$Dirichlet boundary values$: g_0(t)=q_1(x=0,t),\; h_0(t)=q_2(x=0,t);\\
$Neumann boundary values$: g_1(t)=q_{1,x}(x=0,t),\; h_1(t)=q_{2,x}(x=0,t).
\end{array}\label{1.5}\eeqa
where $u_0(x)$ and $v_0(x)$ lie in the Schwartz space.

\begin{figure}
\centering
\includegraphics[width=2.2in,height=1.2in]{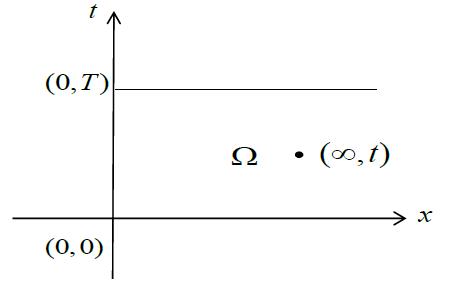}
\caption{The region $\Omega$ in the $(x,t)$ plan}
\label{fig:graph}
\end{figure}

The outline of the present paper is organized as follows. In section 2, we define two sets of eigenfunctions $\{\mu_j\}_1^3$ and $\{M_n\}_1^2$ of Lax pair for spectral analysis and we also get some spectral functions satisfies the so-called global relation in this part. In section 3, we show that $\{q_{1,x}(x,t),q_{2,x}(x,t)\}$ can be expressed in terms of the unique solution of a $4\times 4$ matrix Riemann-Hilbert problem formulated in the complex $\lambda$-plane, and the solution $\{q_{1}(x,t),q_{2}(x,t)\}$ of coupled focusing-defocusing complex short pulse equation can be obtained by integration with respect to $x$. The last section is devoted to giving some conclusions and discussions.

\section{ The spectral analysis}

The coupled focusing-defocusing complex short pulse equation \eqref{1.4} admits the $4\times4$ Lax pair \cite{Yang2017}
 \eqa
\psi_x=U\psi,\,\,\psi_t=V\psi, \label{2.0}\eeqa
with
\eqa
U=\lambda\left(\begin{array}{cc}
I_2&Q_x\\
R_x&-I_2\end{array} \right),
V=\left(\begin{array}{cc}
-\frac{1}{2}\lambda QR-\frac{1}{4\lambda}I_2 & -\frac{1}{2}\lambda QRQ_x+\frac{1}{2}Q\\
-\frac{1}{2}\lambda RQR_x-\frac{1}{2}R & \frac{1}{2}\lambda QR+\frac{1}{4\lambda}I_2
\end{array} \right),
\nn\eeqa
where $I_2$ is a $2\times 2$ identity matrix, and $Q,R$ are $2\times 2$ matrices defined as
\eqa
Q=\lambda\left(\begin{array}{cc}
q_1&q_2\\
-\bar q_2&\bar q_1\end{array} \right),
R=\left(\begin{array}{cc}
\bar q_1 & -q_2\\
-\bar q_2 & \bar q_2\end{array} \right),
\nn\eeqa
Notice that
 \eqa
QR=RQ=(|q_1|^2-|q_2|^2)I_2.\nn\eeqa
Direct computations reveal that the zero-curvature
equation $U_x-V_t+UV-VU=0$ exactly gives
back system \eqref{1.3}.

By replacing $\lambda$ by $i\lambda$, the Lax pair equation \eqref{2.0} can be rewritten as
 \eqa \left\{ \begin{array}{l}
\psi_x=U\psi=(i\lambda\sigma_4+U_0)\psi,\\
\psi_t=V\psi=(\frac{1}{4i\lambda}\sigma_4-\frac{1}{2}i\lambda V_1+\frac{1}{2}V_0)\psi,
\end{array} \right. \label{2.1}\eeqa
where $\sigma_4=diag\{1,1,-1,-1\}$ is a $4\times4$ matrix, $\psi=\psi(x,t,\lambda)$ is a $4\times4$ matrix-valued or a $4\times1$ column vector-valued spectral function, the $4\times4$ matrix-valued functions $U_0,V_0$ and $V_1$ are defined by
\eqa\begin{array}{l}
U_0(x,t)=\left(\begin{array}{cccc}
0&0& q_{1,x}& q_{2,x}\\
0&0& \bar q_{2,x}& \bar q_{1,x}\\
\bar q_{1,x}&-\bar q_{2,x}0&0\\
-\bar q_{2,x}&\bar q_{1,x}&0&0\end{array} \right),
V_0(x,t)=\left(\begin{array}{cccc}
0&0& q_{1}& q_{2}\\
0&0& \bar q_{2}& \bar q_{1}\\
-\bar q_{1}&\bar q_{2}&0&0\\
\bar q_{2}&-\bar q_{1}&0&0\end{array} \right),\\
V_1(x,t)=(|q_1|^2-|q_2|^2)\left(\begin{array}{cccc}
1&0& q_{1,x}& q_{2,x}\\
0&1&\bar q_{1,x}&\bar q_{1,x}\\
\bar q_{1,x}&-q_{2,x}&-1&0\\
-\bar q_{2,x}& q_{1,x}&0&-1\end{array} \right).
\end{array}\label{2.2}\eeqa

\subsection{The closed one-form}

We find that Lax pair Eq.\eqref{2.1} can be rewritten as
\eqa \left\{ \begin{array}{l}
\psi_x-i\lambda\sigma_4\psi=U_1(x,t)\psi,\\
\psi_t-\frac{1}{4i\lambda}\sigma_4\psi=U_2(x,t,\lambda)\psi,
\end{array} \right. \label{2.3}\eeqa
where
 \beq U_1(x,t)=U_0(x,t),\quad U_2(x,t,\lambda)=-\frac{1}{2}i\lambda V_1+\frac{1}{2}V_0.\label{2.4}\eeq
Introduce a new eigenfunction $\mu(x,t,\lambda)$ is defined by the transform
\beq\psi(x,t,\lambda)=\mu(x,t,\lambda)e^{i\lambda\sigma_4 x+\frac{1}{4i\lambda}\sigma_4 t},\label{2.5}\eeq
then the Lax pair Eq.\eqref{2.3} becomes
\eqa \left\{ \begin{array}{l}
\mu_x-i\lambda[\sigma_4,\mu]=U_1(x,t)\mu,\\
\mu_t-\frac{1}{4i\lambda}[\sigma_4,\mu]=U_2(x,t,\lambda)\mu,
\end{array} \right. \label{2.6}\eeqa
and Eq.\eqref{2.6} leads to a full derivative form
\beq d(e^{-i\lambda\hat\sigma_4 x-\frac{1}{4i\lambda}\hat\sigma_4 t}\mu(x,t,\lambda))=W(x,t,\lambda), \label{2.7}\eeq
where the closed one-form $W(x,t,\lambda)$ defined by
\beq W(x,t,\lambda)=e^{-(i\lambda x+\frac{1}{4i\lambda}t)\hat\sigma_4}(U_1(x,t)dx+U_2(x,t,\lambda)dt)\mu, \label{2.8}\eeq
and $\hat\sigma_4$ represents a matrix operator acting on $4\times4$ matrix $X$ by $\hat\sigma_4 X=[\sigma_4,X]$ and by
$e^{x\hat\sigma_4}X=e^{x\sigma_4}Xe^{-x\sigma_4}$ (see \textbf{Lemma 2.6}).

\subsection{ The basic eigenfunction $\mu_j$'s}

We assume that $\{q_1(x,t),q_2(x,t)\}$ is a sufficiently smooth function in the half-line region $\Omega=\{0<x<\infty,0<t<T\}$, and decays sufficiently when $x\rightarrow\infty.$ $\{\mu_j(x,t,\lambda)\}_1^3$ are the $4\times4$ matrix valued functions, based on the Volterra integral equation, we can define the three eigenfunctions $\{\mu_j(x,t,\lambda)\}_1^3$ of Eq.\eqref{2.6} by
\beq \mu_j(x,t,\lambda)=\mathbb{I}+\int_{\gamma_j}e^{(i\lambda x+\frac{1}{4i\lambda}t)\hat\sigma_4}W_j(\xi,\tau,\lambda),\quad j=1,2,3,\label{2.9}\eeq
where $\mathbb{I}=diag\{1,1,1,1\}$ is a $4\times4$ unit matrix, $W_j$ is determined Eq.\eqref{2.8}, it is only used $\mu_j$ in place of $\mu$, and the contours $\{\gamma_j\}_1^3$ are smooth curve from $(x_j,t_j)$ to $(x,t)$, and $(x_1,t_1)=(0,T), (x_2,t_2)=(0,0), (x_3,t_3)=(\infty,t)$ (see Figure 2).

Thus, for the point $(\xi,\tau)$ on the each contour, we have that the following inequalities hold true
\eqa \begin{array}{l}
\gamma_1:\quad x-\xi\geq 0,\quad t-\tau\leq 0;\\
\gamma_2:\quad x-\xi\geq 0,\quad t-\tau\geq 0;\\
\gamma_3:\quad x-\xi\leq 0,\quad t-\tau = 0.
\end{array}\label{2.10}\eeqa

\begin{figure}
\centering
\includegraphics[width=3.8in,height=1.1in]{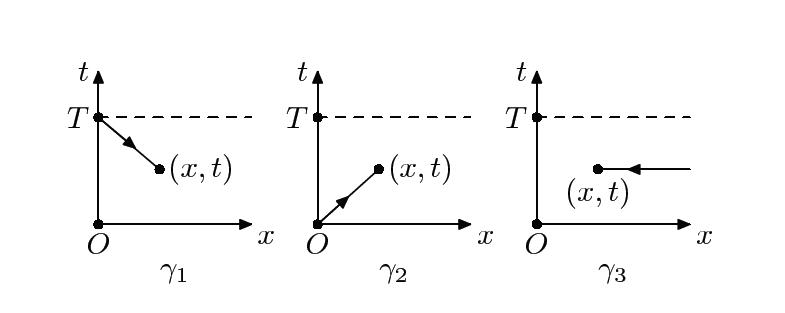}
\caption{The three contours $\gamma_1,\gamma_2,\gamma_3$ in the $(x,t)$-domaint}
\label{fig:graph}
\end{figure}

Since the one-form $W_j$ is closed, thus $\mu_j$ is independent of the path of integration. If we choose the paths of integration to be parallel to the $x$ and $t$ axes, then the integral Eq.\eqref{2.9} becomes $(j=1,2,3)$
\eqa &&\mu_j(x,t,\lambda)=\mathbb{I}+\int_{x_j}^xe^{i\lambda (x-\xi)\hat\sigma_4}(V_1\mu_j)(\xi,t,\lambda)d\xi\nn\\&&
\qquad\qquad\qquad +e^{i\lambda (x-x_j)\hat\sigma_4}\int_{t_j}^te^{\frac{1}{4i\lambda} (t-\tau)\hat\sigma_4}(V_2\mu_j)(x_j,\tau,\lambda)d\tau,\label{2.11}\eeqa
Let ${[\mu_j]}_k$ denote the $k$-th column vector of $\mu_j$, Eq.\eqref{2.10} implies that the first, second, third and fourth columns of the matrices equation\eqref{2.9} contain the following exponential term
\eqa\begin{array}{l}
{[\mu_j]}_1:\, e^{-2i\lambda(x-\xi)-\frac{1}{2i\lambda}(t-\tau)},\, e^{-2i\lambda(x-\xi)-\frac{1}{2i\lambda}(t-\tau)},\\
{[\mu_j]}_2:\, e^{-2i\lambda(x-\xi)-\frac{1}{2i\lambda}(t-\tau)},\, e^{-2i\lambda(x-\xi)-\frac{1}{2i\lambda}(t-\tau)},\\
{[\mu_j]}_3:\, e^{2i\lambda(x-\xi)+\frac{1}{2i\lambda}(t-\tau)},\, e^{2i\lambda(x-\xi)+\frac{1}{2i\lambda}(t-\tau)},\\
{[\mu_j]}_4:\, e^{2i\lambda(x-\xi)+\frac{1}{2i\lambda}(t-\tau)},\, e^{2i\lambda(x-\xi)+\frac{1}{2i\lambda}(t-\tau)}.
\end{array}\label{2.12}\eeqa

Thus, we can show that the eigenfunctions $\{\mu_j(x,t,\lambda)\}_1^3$ are bounded and analytic for $\lambda\in\mathbb{C}$ such that $\lambda$ belongs to
\eqa\begin{array}{l}
 \mu_1 $ is bounded and analytic for $ \lambda\in\emptyset,\\
 \mu_2 $ is bounded and analytic for $ \lambda\in(D_2,D_2,D_1,D_1),\\
 \mu_3 $ is bounded and analytic for $ \lambda\in(D_1,D_1,D_2,D_2),\\
\end{array}\label{2.13}\eeqa
where $D_1,D_2$ denote up-half plane and low-half plane, respectively, pairwisely disjoint subsets of the Riemann $\lambda$-plane shown in figure 3.

\begin{figure}
\centering
\includegraphics[width=2.8in,height=1.0in]{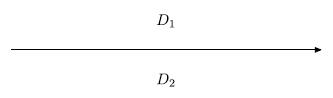}
\caption{The sets $D_n,n=1,2,$ which decompose the complex $\lambda-$plane}
\label{fig:graph}
\end{figure}

And these sets $\{D_n\}_1^2$ have the following properties:
\eqa\begin{array}{l}
D_1=\{\lambda\in\mathbb{C}|\mathrm{Re}l_1=\mathrm{Re}l_2<\mathrm{Re}l_3=\mathrm{Re}l_4,\quad \mathrm{Re}z_1=\mathrm{Re}z_2<\mathrm{Re}z_3=\mathrm{Re}z_4\},\\
D_2=\{\lambda\in\mathbb{C}|\mathrm{Re}l_1=\mathrm{Re}l_2>\mathrm{Re}l_3=\mathrm{Re}l_4,\quad \mathrm{Re}z_1=\mathrm{Re}z_2>\mathrm{Re}z_3=\mathrm{Re}z_4\},
\end{array}\label{2.14}\eeqa
where $l_i(\lambda)$ and $z_i(\lambda)$ are the diagonal elements of the $4\times4$ matrix $i\lambda\sigma_4$ and $\frac{1}{4i\lambda}\sigma_4$.

We note that $\mu_1(x,t,\lambda)$ and $\mu_2(x,t,\lambda)$ are entire functions of $\lambda$. Moreover, in their
corresponding regions of boundedness
\beq \mu_j(x,t,\lambda)=\mathbb{I}+O(\frac{1}{\lambda}),\,\lambda\rightarrow\infty,\,j=1,2,3.\label{2.15}\eeq

\subsection{The symmetry of eigenfunctions}

For the convenience, we write a $4\times4$ matrix $X=(X_{ij})_{4\times4}$ as
\eqa\begin{array}{c}
X=\left(\begin{array}{cc}
\tilde{X}_{11} & \tilde{X}_{12} \\
\tilde{X}_{21} & \tilde{X}_{22}
\end{array}\right),\quad
\tilde{X}_{11}=\left(\begin{array}{cc}
X_{11} & X_{12} \\
X_{21} & X_{22}
\end{array}\right),\quad
\tilde{X}_{12}=\left(\begin{array}{cc}
X_{13} & X_{14} \\
X_{23} & X_{24}
\end{array}\right),\\
\tilde{X}_{21}=\left(\begin{array}{cc}
X_{31} & X_{32} \\
X_{41} & X_{42}
\end{array}\right),\quad
\tilde{X}_{22}=\left(\begin{array}{cc}
X_{33} & X_{34} \\
X_{43} & X_{44}
\end{array}\right).
\end{array}\label{2.16}\eeqa
As $U(x,t,\lambda)=i\lambda\sigma_4+U_0, V(x,t,\lambda)=\frac{1}{4i\lambda}\sigma_4-\frac{1}{2}i\lambda V_1+\frac{1}{2}V_0$. Then the symmetry properties of $U(x,t,\lambda)$ and $ V(x,t,\lambda)$ imply that the eigenfunction $\mu(x,t,\lambda)$ have the symmetries
\beq
(\tilde{\mu}(x,t,\lambda))_{11}=P^{\beta}{\overline{(\tilde{\mu}(x,t,\bar{\lambda}))}}_{22}P^{\beta},\,
(\tilde{\mu}(x,t,\lambda))_{12}=\alpha{\overline{(\tilde{\mu}(x,t,\bar{\lambda}))}}_{21}^T,
\label{2.17}\eeq
where $P^\beta=diag(1,\beta)$ and $\beta^2=1$.

Since
\beq
P_\pm^\alpha{\overline{(\tilde{U}(x,t,\bar{\lambda}))}}P_\pm^\alpha=-U(x,t,\lambda)^T,\,
P_\pm^\alpha{\overline{(\tilde{V}(x,t,\bar{\lambda}))}}P_\pm^\alpha=-V(x,t,\lambda)^T,
\label{2.18}\eeq
where $P_\pm^\alpha=diag(\pm\alpha,\pm\alpha,\mp1,\mp1)$ and $\alpha^2=1$.

According to Eq.\eqref{2.22} (see the similar proof in Ref. \cite{Fokas2005}), we know that the eigenfunction $\psi(x,t,\lambda)$ of the
Lax pair \eqref{2.3} and $\mu(x,t,\lambda)$ of the Lax pair \eqref{2.6} are of the same symmetric relation
\beq
\psi^{-1}(x,t,\lambda)=P_{\pm}^{\alpha}{\overline{(\psi(x,t,\bar{\lambda}))}}^TP_{\pm}^{\alpha},\,
\mu^{-1}(x,t,\lambda)=P_{\pm}^{\alpha}{\overline{(\mu(x,t,\bar{\lambda}))}}P_{\pm}^{\alpha},
\label{2.19}\eeq

Moreover, In the domains where $\mu(x,t,\lambda)$  is bounded, we have
\beq \mu(x,t,\lambda)=\mathbb{I}+O(\frac{1}{\lambda}),\,\lambda\rightarrow\infty,\label{2.20}\eeq
and det$[\mu(x,t,\lambda)]=1$ since $tr(U(x,t,\lambda))=tr(V(x,t,\lambda)=0$.

\subsection{ The adjugated eigenfunction}

We also need to consider the bounded and analytical properties of the minors of the matrices $\{\mu_j(x,t,\lambda)\}_1^3$. We recall that the cofactor matrix $B^A$ of a $4\times4$ matrix $B$ is defined by
\beq B^A=\left(\begin{array}{cccc}
m_{11}(B)&-m_{12}(B)&m_{13}(B)&-m_{14}(B)\\
-m_{21}(B)&m_{22}(B)&-m_{23}(B)&m_{24}(B)\\
m_{31}(B)&-m_{32}(B)&m_{33}(B)&-m_{34}(B)\\
-m_{41}(B)&m_{42}(B)&-m_{43}(B)&m_{44}(B) \end{array}
\right),\label{2.21}\eeq
where $m_{ij}(B)$ denote the $(ij)$th minor of $B$ and $(B^A)^TB=\mathrm{adj}(B)B=\mathrm{det}B$.

It follows from Eq.\eqref{2.6} that be shown that the matrix-valued functions $\mu^A$'s satisfies the Lax pair
\eqa \left\{ \begin{array}{l}
\mu_x^A+i\lambda[\sigma_4,\mu^A]=-V_1^T\mu^A,\\
\mu_t^A+\frac{1}{4i\lambda}[\sigma_4,\mu^A]=-V_2^T\mu^A,
\end{array}\right.\label{2.22}\eeqa
where the superscript $T$ denotes a matrix transpose. Then the eigenfunctions $\{\mu_j^A(x,t,\lambda)\}_1^3$ are solutions solutions can be expressed as
\eqa &&\mu_j^A(x,t,\lambda)=\mathbb{I}-\int_{x_j}^xe^{-i\lambda (x-\xi)\hat\sigma_4}(V_1\mu_j^A)(\xi,t,\lambda)d\xi\nn\\&&
\qquad\qquad\qquad -e^{-i\lambda (x-x_j)\hat\sigma_4}\int_{t_j}^te^{-\frac{1}{4i\lambda} (t-\tau)\hat\sigma_4}(V_2\mu_j^A)(x_j,\tau,\lambda)d\tau,\label{2.23}\eeqa
by using the Volterra integral equations, respectively.
Thus, we can obtain the adjugated eigenfunction which satisfies the following analytic properties
\eqa\begin{array}{l}
 \mu_1^A $ is bounded and analytic for $ \lambda\in\emptyset,\\
 \mu_2^A $ is bounded and analytic for $ \lambda\in(D_1,D_1,D_2,D_2),\\
 \mu_3^A $ is bounded and analytic for $ \lambda\in(D_2,D_2,D_1,D_1).
\end{array}\label{2.24}\eeqa

\subsection{ The spectral functions and the global relation }

We also define the $4\times4$ matrix value spectral function $s(\lambda),S(\lambda)$ and $\mathfrak{S}(\lambda)$ as follows
\eqa\begin{array}{l}
\mu_3(x,t,\lambda)=\mu_2(x,t,\lambda)e^{(i\lambda x+\frac{1}{4i\lambda}t)\hat\sigma_4} s(\lambda),\\
\mu_1(x,t,\lambda)=\mu_2(x,t,\lambda)e^{(i\lambda x+\frac{1}{4i\lambda}t)\hat\sigma_4} S(\lambda),\\
\mu_3(x,t,\lambda)=\mu_1(x,t,\lambda)e^{(i\lambda x+\frac{1}{4i\lambda}t)\hat\sigma_4} \mathfrak{S}(\lambda),
\end{array}\label{2.25}\eeqa
as $\mu_2(0,0,\lambda)=\mathbb{I}$, we obtain
\eqa\begin{array}{l}
s(\lambda)=\mu_3(0,0,\lambda),\\
S(\lambda)=\mu_1(0,0,\lambda)=e^{-\frac{1}{4i\lambda}T\hat\sigma_4}\mu_2^{-1}(0,T,\lambda),\\
\mathfrak{S}(\lambda)=\mu_1^{-1}(0,0,\lambda)\mu_3(0,0,\lambda)=S^{-1}(\lambda)s(\lambda)=e^{-\frac{1}{4i\lambda}T\hat\sigma_4}\mu_3^{-1}(0,T,\lambda).
\end{array}\label{2.26}\eeqa

\begin{figure}
\centering
\includegraphics[width=2.2in,height=1.2in]{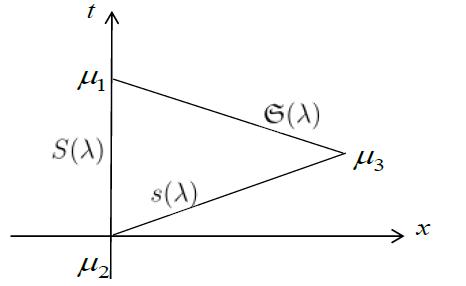}
\caption{The relations among the dependent eigenfunctions $\mu_j(x,t,k), j=1,2,3$.}
\label{fig:graph}
\end{figure}

These relations among $\mu_j$ are displayed in Figure 4. Thus these three functions $s(\lambda),s(\lambda)$ and $\mathfrak{S}(\lambda)$ are
dependent such that we only consider two of them, e.g. $s(\lambda)$ and $S(\lambda)$

According to the definition \eqref{2.11} of $\mu_j$, Eq.\eqref{2.26} implies that
\eqa\begin{array}{l}
s(\lambda)=\mathbb{I}-\int_{0}^{\infty}e^{-i\lambda\xi\hat\sigma_4}(U_1\mu_3)(\xi,0,\lambda)d\xi,\\
S(\lambda)=\mathbb{I}-\int_{0}^{T}e^{-\frac{1}{4i\lambda}\tau\hat\sigma_4}(U_2\mu_1)(0,\tau,\lambda)d\tau\\
\qquad=[\mathbb{I}+\int_{0}^{T}e^{-\frac{1}{4i\lambda}\tau\hat\sigma_4}(U_2\mu_2)(0,\tau,\lambda)d\tau]^{-1},
\end{array}\label{2.27}\eeqa
where $\mu_j(0,t,\lambda)(j=1,2)$ and $\mu_3(x,t,\lambda), 0<x<\infty, 0<t<T$ satisfy the Volterra integral equations
\eqa\begin{array}{l}
\mu_1(0,t,\lambda)=\mathbb{I}-\int_{t}^{T}e^{\frac{1}{4i\lambda}(t-\tau)\hat\sigma_4}(U_2\mu_1)(0,\tau,\lambda)d\tau,\\
\mu_2(0,t,\lambda)=\mathbb{I}+\int_{0}^{T}e^{\frac{1}{4i\lambda}(t-\tau)\hat\sigma_4}(U_2\mu_2)(0,\tau,\lambda)d\tau,\\
\mu_3(x,0,\lambda)=\mathbb{I}-\int_{x}^{\infty}e^{i\lambda(x-\xi)\hat\sigma_4}(U_1\mu_3)(\xi,0,\lambda)d\xi.
\end{array}\label{2.28}\eeqa
Thus, it follows from Eqs.\eqref{2.27} and \eqref{2.28} that $s(\lambda)$ and $S(\lambda)$ are determined by $U(x,0,\lambda)$ and $V(0,t,\mu)$, that is to say, determined by the initial data $u_0(x),v_0(x)$ and the boundary data $g_0(t),h_0(t)$, $g_1(t),h_1(t)$.

Indeed, the eigenfunctions $\mu_3(x,0,\lambda)$ and $\mu_j(0,t,\lambda),j=1,2$ satisfies the $x$-part and $t$-part of the Lax pair \eqref{2.6} at $t=0$ and $x=0$, respectively. Then, we have
\eqa
x-part: \left\{\begin{array}{l}
\mu_x(x,0,\lambda)-i\lambda[\sigma_4,\mu(x,0,\lambda)]=U(x,t=0)\mu(x,0,\lambda),\\
\lim_{x\rightarrow\infty}\mu(x,0,\lambda)=\mathbb{I},\,0<x<\infty,
\end{array}\right.\label{2.29}\eeqa
\eqa
t-part: \left\{\begin{array}{l}
\mu_t(0,t,\lambda)-\frac{1}{4i\lambda}[\sigma_4,\mu(0,t,\lambda)]=V(x=0,t)\mu(0,t,\lambda),\,0<t<T,\\
\lim_{t\rightarrow 0}\mu(0,t,\lambda)=\mu(0,0,\lambda)=\mathbb{I},\\
\lim_{t\rightarrow T}\mu(0,t,\lambda)=\mu(0,T,\lambda)=\mathbb{I}.
\end{array}\right.\label{2.30}\eeqa

Moreover, from the properties of $\{\mu_j\}_1^3$ and $\{\mu_j^A\}_1^3$, we can obtain that $s(\lambda),S(\lambda),s^A(\lambda)$ and $S^A(\lambda)$ have the following bounded properties
\eqa\begin{array}{l}
 s(\lambda) $ is bounded for $ \lambda\in(D_1, D_1, D_2, D_2,),\\
 S(\lambda)$ is bounded for $ \lambda\in\emptyset,\\
 s^A(\lambda)$ is bounded for $ \lambda\in(D_2, D_2, D_1, D_1),\\
 S^A(\lambda)$ is bounded for $ \lambda\in\emptyset.
\end{array}\label{2.31}\eeqa

The spectral functions $S(\lambda)$ and $s(\lambda)$ are not independent which is of important relationship each other. In fact, from Eq.\eqref{2.26}, we have
\beq
\mu_3(x,t,\lambda)=\mu_1(x,t,\lambda)e^{(i\lambda x+\frac{1}{4i\lambda}t)\hat\sigma_4} S^{-1}(\lambda)s(\lambda),
\label{2.32}\eeq
as $\mu_1(0,t,\lambda)=\mathbb{I}$, when $(x,t)=(0,T)$. We can evaluate the following relationship which is the global relation as follows
\beq S^{-1}(\lambda)s(\lambda)=e^{-2i\lambda^2T\hat\sigma_4}c(T,\lambda)=e^{-\frac{1}{4i\lambda}T\hat\sigma_4}\mu_3(0,T,\lambda),
\label{2.33}\eeq
where $\mu_3(0,t,\lambda)$ satisfy the Volterra integral equation
\eqa
\mu_3(0,t,\lambda)=\mathbb{I}-\int_{0}^{\infty}e^{-i\lambda\xi\hat\sigma_4}(V_1\mu_3)(\xi,T,\lambda)d{\xi},\,
0<t<T,\,\lambda\in(D_1, D_1, D_2, D_2,).\label{2.34}\eeqa

\subsection{The definition of matrix-valued functions $M_n$'s}

For each $n=1,2$, the solution $M_n(x,t,\lambda)$ of Eq.\eqref{2.6} is defined by the following integral equation
\beq (M_n(x,t,\lambda))_{ij}=\delta_{ij}+\int_{\gamma_{ij}^n}(e^{(i\lambda x+\frac{1}{4i\lambda}t)\hat\sigma_4}W_n(\xi,\tau,\lambda))_{ij},
\quad i,j=1,2,3,4,\label{2.35}\eeq
where $W_n(x,t,\lambda)$ is given by Eq.\eqref{2.8}, it is only used $M_n$ in place of $\mu$, and the contours $ \gamma_{ij}^n (n=1,2;i,j=1,2,3)$ are defined as follows
 \eqa \gamma_{ij}^n=\left\{\begin{array}{l}
 \gamma_1\quad if\quad Rel_i(\lambda)<Rel_j(\lambda) \quad and \quad Rez_i(\lambda)\geq Rez_j(\lambda),\\
 \gamma_2\quad if\quad Rel_i(\lambda)<Rel_j(\lambda) \quad and \quad Rez_i(\lambda)<Rez_j(\lambda), for\lambda\in D_n\\
 \gamma_3\quad if \quad Rel_i(\lambda)\geq Rel_j(\lambda).
\end{array}\right. \label{2.36}\eeqa
According to the definition of $\gamma^n$, we have
\eqa \begin{array}{l}
\gamma^1=\left(\begin{array}{cccc}
\gamma_3&\gamma_3&\gamma_2&\gamma_2\\
\gamma_3&\gamma_3&\gamma_2&\gamma_2\\
\gamma_3&\gamma_3&\gamma_3&\gamma_3\\
\gamma_3&\gamma_3&\gamma_3&\gamma_3\end{array} \right),\,
\gamma^2=\left(\begin{array}{cccc}
\gamma_3&\gamma_3&\gamma_3&\gamma_3\\
\gamma_3&\gamma_3&\gamma_3&\gamma_3\\
\gamma_2&\gamma_2&\gamma_3&\gamma_3\\
\gamma_2&\gamma_2&\gamma_3&\gamma_3\end{array} \right).
\end{array}\label{2.37}\eeqa

Next, the following proposition guarantees that the previous definition of $M_n$ has properties, namely, $M_n$ can be represented as a RH problem.

\begin{prop}
For each $n=1,2$ and $\lambda\in D_n$, the function $M_n(x,t,\lambda)$ is defined well by Eq.\eqref{2.35}. For any identified point $(x,t)$, $M_n$
is bounded and analytical as a function of $\lambda\in D_n$ away from a possible discrete set of singularities $\{\lambda_j\}$ at which the Fredholm determinant vanishes. Moreover, $M_n$ admits a bounded and continuous extension to $\mathrm{Re} \lambda$-axis and
\beq M_n(x,t,\lambda)=\mathbb{I}+O(\frac{1}{\lambda}).\label{2.38}\eeq
\end{prop}

Proof: The associated bounded and analytical properties have been established in Appendix B in \cite{Lenells2012}. Substituting the following expansion
\beq M=M_0+\frac{M^{(1)}}{\lambda}+\frac{M^{(2)}}{\lambda^2}+\cdots\quad\quad\lambda\rightarrow\infty,\label{2.39}\eeq
into the Lax pair Eq.\eqref{2.6} and comparing the coefficients of the same order of $\lambda$, we can obtain Eq.\eqref{2.39}.

\begin{rem}
So far, we define two sets of eigenfunctions $\{\mu_j\}_1^3$ and $\{M_n\}_1^2$. The Fokas method in \cite{Fokas1997} analyzed the $2\times 2$ Lax pair related to two kinds of eigenfunctions $\mu_j$, which is used for spectral analysis, and the other eigenfunction is used to be shown RH problem, our definition on $M_n$ is similar to the latter eigenfunction.
\end{rem}

\subsection{ The jump matrix and computations}

The new spectral functions $S_n(\lambda)(n=1,2)$ are defined by
\beq S_n(\lambda)=M_n(0,0,\lambda),\, \lambda\in D_n,n=1,2.\label{2.40}\eeq
Let $M(x,t,\lambda)$ be a sectionally analytical continuous function in Riemann $\lambda$-sphere which equals $M_n(x,t,\lambda)$ for
$\lambda\in D_n$. Then $M(x,t,\lambda)$ satisfies the following jump conditions
\beq M_1(x,t,\lambda)=M_2(x,t,\lambda)J(x,t,\lambda),\, \lambda\in \mathbb{R},\label{2.41}\eeq
where
\beq J(x,t,\lambda)=e^{(i\lambda x+\frac{1}{4i\lambda}t)\hat\sigma_4}[S_2^{-1}(\lambda)S_1(\lambda)].\label{2.42}\eeq

\begin{rem}
As the integral equation \eqref{2.35} defined by $M_n(0,0,\lambda)$ involves only along the initial half-line $\{0<x<\infty,t=0\}$
and along the boundary $\{x=0,0<t<T\}$ , so $S_n$'s can only be determined by the initial data
and boundary data, therefore, equation \eqref{2.42} represents a jump condition of RH problem.
In the absence of singularity, the solution $q_1(x,t),q_2(x,t)$ of the equation can be reconstructed from the initial data and boundary values data,
but if the $M_n$ have pole singularities at some point $\{\lambda_j\},\lambda_j\in\mathbb{C}$, the RH problem
should be included the residue condition in these points, so in order to determine the correct residue condition,
we need to introduce three eigenfunctions $\{\mu_j(x,t,\lambda)\}_1^3$ in addition to the $M_n$'s.
\end{rem}

\begin{prop}
The matrix-valued functions $S_n(x,t,\lambda),(n=1,2)$ defined by
\beq M_n(x,t,\lambda)=\mu_2(x,t,\lambda)e^{(i\lambda x+\frac{1}{4i\lambda}t)\hat\sigma_4}S_n(\lambda),\, \lambda\in D_n.\label{2.43}\eeq
can be expressed with $s(\lambda)$ and $S(\lambda)$ elements as follows
\eqa \begin{array}{l}
S_1(\lambda)=\left(\begin{array}{cccc}
s_{11}&s_{12}&0&0\\
s_{21}&s_{22}&0&0\\
s_{31}&s_{32}&\frac{m_{44}(s)}{n_{11,22}(s)}&\frac{m_{43}(s)}{n_{11,22}(s)}\\
s_{41}&s_{42}&\frac{m_{34}(s)}{n_{11,22}(s)}&\frac{m_{33}(s)}{n_{11,22}(s)}
\end{array} \right),\,
S_2(\lambda)=\left(\begin{array}{cccc}
\frac{m_{22}(s)}{n_{33,44}(s)}&\frac{m_{21}(s)}{n_{33,44}(s)}&s_{13}&s_{14}\\
\frac{m_{12}(s)}{n_{33,44}(s)}&\frac{m_{11}(s)}{n_{33,44}(s)}&s_{23}&s_{24}\\
0&0&s_{33}&s_{34}\\
0&0&s_{43}&s_{44}
\end{array} \right).
\end{array}\label{2.44}\eeqa
where $n_{i1j1,i2j2}(X)$ denotes the determinant of the sub-matrix generated by taking the cross elements of $i_{1,2}$ th
rows and $j_{1,2}$ th columns of the $4\times4$ matrix $X$. that is to say
$$
n_{i1j1,i2j2}(X)=\left|\begin{array}{cc}
X_{i1j1} & X_{i1j2}\\
X_{i2j1} & X_{i2j2}\end{array} \right|,
$$
\end{prop}

Proof: We set that $\gamma_3^{X_0}$ is a contour when $(X_0,0)\rightarrow (x,t)$ in the $(x,t)$-plane, here $X_0$ is a constant and $X_0>0$, for $j=3$, we introduce $\mu_3(x,t,\lambda;X_0)$ as the solution of Eq.\eqref{2.9} with the contour $\gamma_3$ replaced by $\gamma_3^{X_0}$. Similarly, we define $M_n(x,t,\lambda;X_0)$ as the solution of Eq.\eqref{2.35} with $\gamma_3$ replaced by $\gamma_3^{X_0}$. Then, by simple calculation, we can use $S(\lambda)$ and $s(\lambda;X_0)=\mu_3(0,0,\lambda;X_0)$ to derive the expression of $S_n(\lambda,X_0)=M_n(0,0,\lambda;X_0)$ and the Eq.\eqref{2.44} will be obtained by taking the limit $X_0\rightarrow\infty$.

Firstly, we have the following relations:
\eqa \begin{array}{l}
M_n(x,t,\lambda;X_0)=\mu_1(x,t,\lambda)e^{(i\lambda x+\frac{1}{4i\lambda}t)\hat\sigma_4} R_n(\lambda;X_0),\\
M_n(x,t,\lambda;X_0)=\mu_2(x,t,\lambda)e^{(i\lambda x+\frac{1}{4i\lambda}t)\hat\sigma_4} S_n(\lambda;X_0),\\
M_n(x,t,\lambda;X_0)=\mu_3(x,t,\lambda;X_0)e^{(i\lambda x+\frac{1}{4i\lambda}t)\hat\sigma_4} T_n(\lambda;X_0).
\end{array} \label{2.47}\eeqa

Secondly, we can get the definition of $R_n(\lambda;X_0)$ and $T_n(\lambda;X_0)$ as follows
\eqa \begin{array}{l}
R_n(\lambda;X_0)=e^{-\frac{1}{4i\lambda}T\hat\sigma} M_n(0,T,\lambda;X_0),\\
T_n(\lambda;X_0)=e^{-i\lambda X_0\hat\sigma} M_n(X_0,0,\lambda;X_0),
\end{array} \label{2.48}\eeqa
then equation \eqref{2.47} mean that
\eqa \begin{array}{l}
s(\lambda;X_0)=S_n(\lambda;X_0)T_n^{-1}(\lambda;X_0),\\
S(\lambda;X_0)=S_n(\lambda;X_0)R_n^{-1}(\lambda;X_0).
\end{array} \label{2.49}\eeqa

These equations constitute the matrix decomposition problem of $\{s,S\}$ by use $\{R_n,S_n,T_n\}$. In fact, by the definition of the integral equation \eqref{2.35} and $\{R_n,S_n,T_n\}$, we obtain
\eqa \left \{\begin{array}{l}
(R_n(\lambda;X_0))_{ij}=0\quad if \quad \gamma_{ij}^n=\gamma_1,\\
(S_n(\lambda;X_0))_{ij}=0\quad if \quad \gamma_{ij}^n=\gamma_2,\\
(T_n(\lambda;X_0))_{ij}=\delta_{ij}\quad if \quad \gamma_{ij}^n=\gamma_3.
\end{array}\right. \label{2.50}\eeqa

Thus equation \eqref{2.48} contains 32 scalar equations for 32 unknowns. The exact solution of these system can be obtained by solving the algebraic system. In this way, we can get a similar $\{S_n(\lambda),s(\lambda)\}$ as in Eq.\eqref{2.44} which just that $\{S_n(\lambda),s(\lambda)\}$ replaces by $\{S_n(\lambda;X_0),s(\lambda;X_0)\}$ in Eq.\eqref{2.44}.

 Finally, taking $X_0\rightarrow\infty$ in this equation, we obtain the Eq.\eqref{2.44}.

\subsection{ The residue conditions}

Because $\mu_2(x,t,\lambda)$ is an entire function, and from Eq.\eqref{2.43} we know that $M(x,t,\lambda)$ only produces singularities in $S_n(\lambda)$ where there are singular points, from the exact expression Eq.\eqref{2.44}, we know that $M(x,t,\lambda)$ may be singular as follows
\begin{itemize}
  \item $[M_1]_3$ and $[M_1]_4$ could have poles in $D_1$ at the zeros of $n_{11,22}(s)(\lambda)$,
  \item $[M_2]_1$ and $[M_2]_2$ could have poles in $D_2$ at the zeros of $n_{33,44}(s)(\lambda)$.
\end{itemize}

We use $\{\lambda_j\}_1^N$ denote the possible zero point above, and assume that these zeros satisfy the following assumptions
\begin{ass}
Suppose that
\begin{itemize}
  \item $n_{11,22}(s)(\lambda)$ admits $n_1$ possible simple zeros in $D_1$ denoted by $\{\lambda_j\}_{1}^{n_1}$,
  \item $n_{33,44}(s)(\lambda)$ admits $N-n_1$ possible simple zeros in $D_3$ denoted by $\{\lambda_j\}_{n_1+1}^{N}$.
\end{itemize}
\end{ass}
And these zeros are each different, moreover assuming that there is no zero on the boundary of $D_n$'s$(n=1,2)$.

\begin{lem}
For a $4\times 4$ matrix $X=(X_{ij})_{4\times 4}, e^{\theta\hat\sigma_4}X $ is given by
$$
e^{\theta\hat\sigma_4}X=e^{\theta\sigma_4}Xe^{-\theta\sigma_4}=\left(\begin{array}{cccc}
X_{11} & X_{12} & X_{13}e^{2\theta} & X_{14}e^{2\theta}\\
X_{21} & X_{22} & X_{23}e^{2\theta} & X_{24}e^{2\theta}\\
X_{31}e^{-2\theta} & X_{32}e^{-2\theta} & X_{33} & X_{34}\\
X_{41}e^{-2\theta} & X_{42}e^{-2\theta} & X_{43} & X_{44}\\\end{array} \right).
$$
We can deduce the residue conditions at these zeros in the following expressions:
\end{lem}

\begin{prop}
Let $\{M_n(x,t,\lambda)\}_1^4$ be the eigenfunctions defined by Eq.\eqref{2.35} and assume that the set $\{\lambda_j\}_1^N$  of singularities are as the above assumption. Then the following residue conditions hold true:
\eqa \begin{array}{l}
\mathrm{Res}_{\lambda=\lambda_j}[M_1(x,t,\lambda)]_k=\frac{m_{4(5-k)}(s)(\lambda_j)s_{42}(\lambda_j)-m_{3(5-k)}(s)(\lambda_j)s_{32}(\lambda_j)}
{\dot{n_{11,22}(s)(\lambda_j)}n_{31,42}(s)(\lambda_j)}[M_1(x,t,\lambda_j)]_3e^{\theta_{13}(\lambda_j)}\\
\qquad\qquad\qquad\qquad\qquad+\frac{m_{3(5-k)}(s)(\lambda_j)s_{31}(\lambda_j)-m_{4(5-k)}(s)(\lambda_j)s_{41}(\lambda_j)}
{\dot{n_{11,22}(s)(\lambda_j)}n_{31,42}(s)(\lambda_j)}[M_1(x,t,\lambda_j)]_4e^{\theta_{13}(\lambda_j)},\\
\qquad\qquad\qquad\qquad\qquad n_1+1\leq j\leq n_2 ;\lambda_j\in D_1, k=1,2.
\end{array}\label{2.52}\eeqa
\eqa \begin{array}{l}
\mathrm{Res}_{\lambda=\lambda_j}[M_2(x,t,\lambda)]_k=\frac{m_{2(5-k)}(s)(\lambda_j)s_{24}(\lambda_j)-m_{1(5-k)}(s)(\lambda_j)s_{14}(\lambda_j)}
{\dot{n_{33,44}(s)(\lambda_j)}n_{13,24}(s)(\lambda_j)}[M_2(x,t,\lambda_j)]_1e^{\theta_{31}(\lambda_j)}\\
\qquad\qquad\qquad\qquad\qquad+\frac{m_{1(5-k)}(s)(\lambda_j)s_{13}(\lambda_j)-m_{2(5-k)}(s)(\lambda_j)s_{23}(\lambda_j)}
{\dot{n_{33,44}(s)(\lambda_j)}n_{13,24}(s)(\lambda_j)}[M_2(x,t,\lambda_j)]_2e^{\theta_{31}(\lambda_j)},\\
\qquad\qquad\qquad\qquad\qquad n_2+1\leq j\leq n_3 ;\lambda_j\in D_2, k=3,4.
\end{array}\label{2.53}\eeqa
where $\dot{f}=\frac{df}{d\lambda}$ and $\theta_{ij}$ defined by
\beq \theta_{ij}(x,t,\lambda)=(l_i-l_j)x-(z_i-z_j)t, \quad i,j=1,2,3,4,\label{2.55}\eeq
thus, we have
\eqa \begin{array}{l}
\theta_{12}=\theta_{21}=\theta_{34}=\theta_{43}=0,\\
\theta_{13}=\theta_{14}=\theta_{23}=\theta_{24}=2i\lambda x+\frac{1}{2i\lambda}t,\\
\theta_{31}=\theta_{41}=\theta_{32}=\theta_{42}=-2i\lambda x-\frac{1}{2i\lambda}t.
\end{array}\nn\eeqa
\end{prop}

Proof: The equation \eqref{2.43} mean that
\beq M_1(x,t,\lambda)=\mu_2(x,t,\lambda)e^{(i\lambda x+\frac{1}{4i\lambda}t)\hat\sigma_4}S_1,\label{2.56a}\eeq
\beq M_2(x,t,\lambda)=\mu_2(x,t,\lambda)e^{(i\lambda x+\frac{1}{4i\lambda}t)\hat\sigma_4}S_2,\label{2.56b}\eeq
In view of the expressions for $S_1$ given in \eqref{2.44}, the four columns of Eq.\eqref{2.56a} read
\eqa [M_1]_1=[\mu_2]_1s_{11}+[\mu_2]_2s_{21}+[\mu_2]_3s_{31}e^{\theta_{31}}+[\mu_2]_4s_{41}e^{\theta_{31}},\label{2.57a}\eeqa
\eqa [M_1]_2=[\mu_2]_1s_{12}+[\mu_2]_2s_{22}+[\mu_2]_3s_{32}e^{\theta_{31}}+[\mu_2]_4s_{42}e^{\theta_{31}},\label{2.58a}\eeqa
\eqa [M_1]_3=[\mu_2]_3\frac{m_{44}(s)}{n_{11,22}(s)}+[\mu_2]_4\frac{m_{34}(s)}{n_{11,22}(s)},\label{2.59a}\eeqa
\eqa [M_1]_4=[\mu_2]_3\frac{m_{43}(s)}{n_{11,22}(s)}+[\mu_2]_4\frac{m_{33}(s)}{n_{11,22}(s)}.\label{2.60a}\eeqa
In view of the expressions for $S_2$ given in \eqref{2.44}, the four columns of Eq.\eqref{2.56b} read
\eqa [M_2]_1=[\mu_2]_1\frac{m_{22}(s)}{n_{33,44}(s)}+[\mu_2]_2\frac{m_{12}(s)}{n_{33,44}(s)},\label{2.57b}\eeqa
\eqa [M_2]_2=[\mu_2]_1\frac{m_{21}(s)}{n_{33,44}(s)}+[\mu_2]_2\frac{m_{11}(s)}{n_{33,44}(s)},\label{2.58b}\eeqa
\eqa [M_2]_3=[\mu_2]_1s_{13}e^{\theta_{13}}+[\mu_2]_2s_{23}e^{\theta_{13}}+[\mu_2]_3s_{33}+[\mu_2]_4s_{43},\label{2.59b}\eeqa
\eqa [M_2]_4=[\mu_2]_1s_{14}e^{\theta_{13}}+[\mu_2]_2s_{24}e^{\theta_{13}}+[\mu_2]_3s_{34}+[\mu_2]_4s_{44}.\label{2.60b}\eeqa

Suppose that $\lambda_j\in D_1$ is a simple zero of $n_{11,22}(s)(\lambda)$. Solving Eqs.\eqref{2.59a} and \eqref{2.60a} for $[\mu_2]_3,[\mu_2]_4$ and substituting the result into Eqs.\eqref{2.57a} and \eqref{2.58a}, we find
\eqa && [M_1]_1=\frac{m_{44}(s)s_{42}-m_{34}(s)s_{32}}{n_{11,22}(s)n_{31,42}(s)}[M_1]_3e^{\theta_{13}}
+\frac{m_{34}(s)s_{31}-m_{44}(s)s_{41}}{n_{11,22}(s)n_{13,24}(s)}[M_1]_4e^{\theta_{13}}\nn\\&&
\qquad\quad+\frac{m_{24}(s)[\mu_2]_1+m_{14}(s)[\mu_2]_2}{n_{31,42}(s)}e^{\theta_{13}},\label{2.61a}\eeqa

\eqa && [M_1]_2=\frac{m_{43}(s)s_{42}-m_{33}(s)s_{32}}{n_{11,22}(s)n_{31,42}(s)}[M_1]_3e^{\theta_{13}}
+\frac{m_{33}(s)s_{31}-m_{43}(s)s_{41}}{n_{11,22}(s)n_{31,42}(s)}[M_1]_4e^{\theta_{13}}\nn\\&&
\qquad\quad+\frac{m_{23}(s)[\mu_2]_1+m_{13}(s)[\mu_2]_2}{n_{31,42}(s)}e^{\theta_{13}},\label{2.62a}\eeqa

Taking the residue of this equations at $\lambda_j$, we find condition Eqs.\eqref{2.61a} and \eqref{2.62a} in the case when $\lambda_j\in D_1$.

In the same way, suppose that $\lambda_j\in D_2$ is a simple zero of $n_{33,44}(s)(\lambda)$. Solving Eqs.\eqref{2.59b} and \eqref{2.60b} for $[\mu_2]_1,[\mu_2]_2$ and substituting the result into Eqs.\eqref{2.57b} and \eqref{2.58b}, we find
\eqa && [M_2]_3=\frac{m_{22}(s)s_{24}-m_{12}(s)s_{14}}{n_{33,44}(s)n_{13,24}(s)}[M_2]_1e^{\theta_{13}}
+\frac{m_{12}(s)s_{13}-m_{22}(s)s_{23}}{n_{33,44}(s)n_{13,24}(s)}[M_2]_2e^{\theta_{13}}\nn\\&&
\qquad\quad+\frac{m_{42}(s)[\mu_2]_3+m_{32}(s)[\mu_2]_4}{n_{13,24}(s)}e^{\theta_{13}},\label{2.61b}\eeqa

\eqa && [M_2]_4=\frac{m_{21}(s)s_{24}-m_{11}(s)s_{14}}{n_{33,44}(s)n_{13,24}(s)}[M_2]_1e^{\theta_{13}}
+\frac{m_{11}(s)s_{13}-m_{21}(s)s_{23}}{n_{33,44}(s)n_{13,24}(s)}[M_2]_2e^{\theta_{13}}\nn\\&&
\qquad\quad+\frac{m_{41}(s)[\mu_2]_3+m_{31}(s)[\mu_2]_4}{n_{13,24}(s)}e^{\theta_{13}},\label{2.62b}\eeqa

Taking the residue of this equations at $\lambda_j$, we find condition Eqs.\eqref{2.61b} and \eqref{2.62b} in the case when $\lambda_j\in D_2$.

\section{ The Riemann-Hilbert problem }

In section 2, we define the sectionally analytical function $M(x,t,\lambda)$ that its satisfies a RH problem which can be formulated in terms of the initial and boundary values of $\{q_1(x,t),q_2(x,t)\}$. For all $(x,t)$, the solution of system \eqref{1.4} can be recovered by solving this RH problem. So we can establish the following theorem.
\begin{thm}
Suppose that $\{q_1(x,t),q_2(x,t)\}$ is solution of system \eqref{1.4} in the half-line domain $\Omega$, and it is sufficient smoothness and decays when $x\rightarrow\infty$. Then the solution $u(x,t)$ and $v(x,t)$ of system \eqref{1.4} can be reconstructed from the initial values $\{u_0(x),v_0(x)\}$ and boundary values $\{g_0(t),h_0(t),g_1(t),h_1(t)\}$ defined as follows
\eqa\begin{array}{l}
$Initial values$: u_0(x)=u(x,t=0),\; v_0(x)=v(x,t=0);\\
$Dirichlet boundary values$: g_0(t)=u(x=0,t),\; h_0(t)=v(x=0,t);\\
$Neumann boundary values$: g_1(t)=u_x(x=0,t),\; h_1(t)=v_x(x=0,t).
\end{array}\label{3.1}\eeqa
Like Eq.\eqref{2.25} using the initial and boundary data to define the spectral functions $s(\lambda)$ and $S(\lambda)$, we can further define the jump matrix $J(x,t,\lambda)$. Assume that the zero points of the $n_{11,22}(s)(\lambda)$ and $n_{33,44}(s)(\lambda)$, just like in \textbf{Assumption 2.5}. We also have the following results
\eqa\begin{array}{l}
q_{1,x}(x,t)=-2i\lim_{\lambda\rightarrow\infty}(\lambda M(x,t,\lambda))_{13},\\
q_{2,x}(x,t)=-2i\lim_{\lambda\rightarrow\infty}(\lambda M(x,t,\lambda))_{14}.
\end{array}\label{3.2}\eeqa
where $M(x,t,\lambda)$ satisfies the following $4\times 4$ matrix RH problem:
\begin{itemize}
  \item $M(x,t,\lambda)$ is a sectionally meromorphic on the Riemann $\lambda$-sphere with jumps across the contours on $\mathrm{Re} \lambda$-axis (see figure 3).
  \item $M(x,t,\lambda)$ satisfies the jump condition with jumps across the contours on $\mathrm{Re} \lambda$-axis
        \beq M_2(x,t,\lambda)=M_1(x,t,\lambda)J(x,t,\lambda),\quad \lambda\in \mathbb{R}.\label{3.3}\eeq
  \item $M(x,t,\lambda)=\mathbb{I}+O(\frac{1}{\lambda}),\quad \lambda\rightarrow\infty.$
  \item The residue condition of $M(x,t,\lambda)$ is showed in \textbf{Proposition 2.7}.
\end{itemize}
\end{thm}

Proof: We can use similar method like ref. \cite{Xu2013} to prove this Theorem. It only remains to prove Eq.\eqref{3.2} and this equation hold true from the large $\lambda$ asymptotic of the eigenfunctions. We omit this proof in here because of the length of this article.

Thus, the solution of the coupled focusing-defocusing complex short pulse equation $\{q_1(x,t),q_2(x,t)\}$ can be obtained by integration with respect to $x$.

\section{Conclusions and discussions}

In this paper, we consider IBV of the coupled focusing-defocusing complex short pulse equation on the half-line. Using the Fokas method for nonlinear evolution equations which taking the form of Lax pair isospectral deformations and whose corresponding continuous spectra Lax operators, assume that the solution $\{q_1(x,t),q_2(x,t)\}$ exists, we show that it can be represented in terms of the solution of a $4\times 4$ matrix RH problem formulated in the plane of the complex spectral parameter $\lambda$. The spectral functions $s(\lambda)$ and $S(\lambda)$ are not independent, but related by a compatibility condition, the so-called global relation. For other integrable equations with high-order matrix Lax pair, can we construct their solution of a matrix RH problem formulated in the plane of the complex spectral parameter $\lambda$ by the similar method? This question will be discussed in our future paper.

\subsection*{Acknowledgements}

This work is partially supported by the National Natural Science Foundation of China under
Grant Nos. 12271008 and 11601055,  Natural Science Research Project of Universities of Anhui Province under Grant No.1408085QA06

\end{document}